# PID Optimization Using Lagrangian Mechanics

Ethan Kou[1] and Majid Moghadam [2,#]
[1] Henry M Gunn High School, Palo Alto, CA, USA
[2] University of California Santa Cruz   [#] Advisor
https://github.com/BubblyBingBong/PID

Creating a simulation of a system enables the tuning of control systems without the need for a physical system. In this paper, we employ Lagrangian Mechanics to derive a set of equations to simulate an inverted pendulum on a cart. The system consists of a freely-rotating rod attached to a cart, with the rod's balance achieved through applying the correct forces to the cart. We manually tune the proportional, integral, and derivative gain coefficients of a Proportional Integral Derivative controller (PID) to balance a rod. To further improve PID performance, we can optimize an objective function to find better gain coefficients.

## 1. INTRODUCTION

### 1.1. PID Controllers

A PID is a simple and computationally inexpensive controller to implement. Other traditional controllers include Bang-Bang (On-Off) controllers and Model Predictive Control (MPC) [1][2]. There are also learning-based controllers that utilize deep learning techniques, such as imitation learning and reinforcement learning [3].

The PID controllers make decisions based solely on the state error without requiring the system model. Let e(t) denote the error in the state as a function of time, where the error is the difference between the target and the actual state. The control action u(t) over time is the following:

$$u(t) = K_p e(t) + K_i \int_0^t e(t)\, dt + K_d e'(t) \qquad (1)$$

where constants Kp, Ki, and Kd are the proportional gain, integral gain, and derivative gain, respectively. These constants must be tuned to maximize PID performance.

### 1.2. System Overview

In team settings, tuning PIDs for the physical system can prove to be exceedingly time-consuming, often hindering progress on other critical tasks [1]. However, by developing a simulation of the physical system based on its physics model, we can significantly streamline the PID tuning procedure, making it more manageable and efficient. In this paper we develop an inverted pendulum on a cart simulation (Figure 1) and tune a PID in two ways to balance the rod: manually and with optimization techniques. Developing more effective PID tuning methods is important because PIDs are used very commonly in real life, such as cruise control on a car or temperature control on a thermostat.

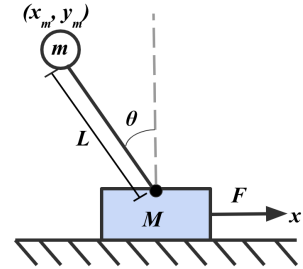

Figure 1. Inverted pendulum on a cart diagram: x = cart position, $\theta$ = rod angle, $x_m$ = ball x position, $y_m$ = ball y position, M = cart mass, m = ball mass, F = net force on cart, and L = rod length. SI units are used.

The assumptions made for this scenario are as follows:
1. The rod is massless
2. The ball is approximated as a point mass.
3. There is no rotational friction acting on the rod.
4. Friction is present between the cart and the floor.

In this paper, we specifically consider e(t) to be the error in the rod angle. To counteract a positive error in the rod angle, a negative force is applied to the cart. Conversely, negative errors result in applying a positive force. Hence, the PID controller is limited to the range of $-\pi/2 < \theta < \pi/2$ because when $\theta$ is out of that range, negative cart forces will increase positive errors, and positive cart forces will decrease negative errors.

## 2. CREATING THE SIMULATION

### 2.1. Simulation Information

The simulation runs with an update time of $\Delta t = 0.001s$. Every update, the position, velocity, and acceleration are calculated. These rules apply for $\theta$ as well:

$$x_{t+1} = x_t + \dot{x}_t \Delta t + \tfrac{1}{2} \ddot{x}_t \Delta t^2 \qquad (2)$$

$$\dot{x}_{t+1} = \dot{x}_t + \ddot{x}_t \Delta t \qquad (3)$$

where the double dot indicates acceleration. (2) and (3) are the basic kinematics equations of physics. Acceleration is calculated using Lagrangian Mechanics. The derivation for acceleration is in the Lagrangian Mechanics section. Although the acceleration is constantly changing, we assume that it only changes once per update, and the system parameters are as follows:

| Parameters | m | M | L | μ | g |
|---|---|---|---|---|---|
| Value | 5 kg | 5 kg | 1 m | 0.3 | 9.8 |

## 2.2. Lagrangian Mechanics

### 1) Setting up the Equations

To calculate the linear acceleration of the cart and angular acceleration of the rod in terms of other variables, we use Lagrangian Mechanics [4]. Unlike Newtonian Mechanics which is based on forces, Lagrangian Mechanics is based on energies. We use Lagrangian Mechanics for this system because calculating the kinetic and potential energy is relatively straightforward. After applying Lagrangian Mechanics, we have the following equations:

$$L = KE - PE \quad (4)$$

$$\frac{d}{dt}\frac{\partial L}{\partial \dot{x}} - \frac{\partial L}{\partial x} = F(t) \quad (5)$$

$$\frac{d}{dt}\frac{\partial L}{\partial \dot{\theta}} - \frac{\partial L}{\partial \theta} = 0 \quad (6)$$

The Lagrangian L is an expression for the difference between the kinetic energy and the potential energy of the entire system. We apply Lagrangian Mechanics to both components of the state, x and $\theta$. The force applied onto a state component is on the right side of the equation. A net force as a function of time F(t) is applied to the cart, and no force is directly applied to the rod angle.

### 2) Calculating the Lagrangian

We first calculate the kinetic energy of the system:

$$KE = \frac{1}{2}M\dot{x}^2 + \frac{1}{2}m(\dot{x}_m^2 + \dot{y}_m^2) \quad (7)$$

$\dot{x}_m^2$ and $\dot{y}_m^2$ can be expressed in terms of x and $\theta$:

$$x_m = x - L\sin\theta \quad (8)$$

$$y_m = L\cos\theta \quad (9)$$

Differentiating both sides of both equations gives

$$\dot{x}_m = \dot{x} - L\dot{\theta}\cos\theta \quad (10)$$

$$\dot{y}_m = -L\dot{\theta}\sin\theta \quad (11)$$

Plugging (10) and (11) into (7) gives

$$KE = \frac{1}{2}M\dot{x}^2 + \frac{1}{2}m\left((\dot{x} - L\dot{\theta}\cos\theta)^2 + (-L\dot{\theta}\sin\theta)^2\right) \quad (12)$$

Expanding and simplifying, we get

$$KE = \frac{1}{2}M\dot{x}^2 + \frac{1}{2}m(\dot{x}^2 - 2\dot{x}L\dot{\theta}\cos\theta$$
$$+ L^2\dot{\theta}^2(\cos\theta)^2 + L^2\dot{\theta}^2(\sin\theta)^2)$$
$$= \frac{1}{2}M\dot{x}^2 + \frac{1}{2}m(\dot{x}^2 - 2\dot{x}L\dot{\theta}\cos\theta + L^2\dot{\theta}^2)$$
$$= \frac{1}{2}M\dot{x}^2 + \frac{1}{2}m\dot{x}^2 - m\dot{x}L\dot{\theta}\cos\theta + \frac{1}{2}mL^2\dot{\theta}^2 \quad (13)$$

Now, we need to find the potential energy of the system. The potential energy of the cart and rod is 0. The potential energy of the ball is the following:

$$PE = mgy_m = mgL\cos\theta \quad (14)$$

Combining (13) and (14), we can calculate the Lagrangian:

$$L = KE - PE = \frac{1}{2}M\dot{x}^2 + \frac{1}{2}m\dot{x}^2$$
$$- m\dot{x}L\dot{\theta}\cos\theta + \frac{1}{2}mL^2\dot{\theta}^2 - mgL\cos\theta \quad (15)$$

### 3) Plugging into Lagrange's Equations

The next step is to plug (15) into (5) and (6). We first substitute into (5).

$$\frac{\partial L}{\partial x} = 0 \quad (16)$$

$$\frac{d}{dt}\left(\frac{\partial L}{\partial \dot{x}}\right) = \frac{d}{dt}((M+m)\dot{x} - mL\dot{\theta}\cos\theta) \quad (17)$$

$$= (M+M)\ddot{x} - mL(\ddot{\theta}\cos\theta - \dot{\theta}^2\sin\theta) \quad (18)$$

$$= (M+m)\ddot{x} - mL\ddot{\theta}\cos\theta + mL\dot{\theta}^2\sin\theta \quad (19)$$

Combining (5), (16), and (19), we get

$$(M+m)\ddot{x} - mL\ddot{\theta}\cos\theta + mL\dot{\theta}^2\sin\theta = F(t) \quad (20)$$

The net force F(t) is calculated from frictional force $\mu(M+m)g$ and the applied force u(t):

$$if\ |\mu(M+m)g| > |u(t)|: F(t) = 0$$
$$else\ if\ u(t) > 0: F(t) = u(t) - \mu(M+m)g$$
$$else: F(t) = u(t) + \mu(M+m)g \quad (21)$$

Then, we substitute (15) into (6):

$$\frac{\partial L}{\partial \theta} = m\dot{x}L\dot{\theta}\sin\theta + mgL\sin\theta \quad (22)$$

$$\frac{d}{dt}\left(\frac{\partial L}{\partial \dot{\theta}}\right) = \frac{d}{dt}(-m\dot{x}L\cos\theta + mL^2\dot{\theta}) \quad (23)$$

$$= -mL\ddot{x}\cos\theta + mL\dot{x}\dot{\theta}\sin\theta + mL^2\ddot{\theta} \quad (24)$$

Combining (6), (22), and (24), we get

$$-mL\ddot{x}\cos\theta + mL\dot{x}\dot{\theta}\sin\theta + mL^2\ddot{\theta}$$
$$- m\dot{x}L\dot{\theta}\sin\theta - mgL\sin\theta = 0 \quad (25)$$

Divide both sides by mL and simplify:

$$-\ddot{x}\cos\theta + \dot{x}\dot{\theta}\sin\theta + L\ddot{\theta}$$

$$-\dot{x}\dot{\theta} \sin \theta - g \sin \theta = 0 \quad (26)$$

$$-\ddot{x} \cos \theta + L\ddot{\theta} - g \sin \theta = 0 \quad (27)$$

*4) Solving for Acceleration*

We can treat (20) and (27) as a system of linear equations in terms of $\ddot{x}$ and $\ddot{\theta}$ to solve for $\ddot{x}$ and $\ddot{\theta}$ in terms of the other variables. These are the coefficients of the linear equation:

$$A = M + m; B = -mL \cos \theta; C = mL\dot{\theta}^2 \sin \theta - F(t)$$

$$D = -\cos \theta; E = L; F = -g \sin \theta \quad (28)$$

These substitutions allow to simplify (20) and (27)

$$A\ddot{x} + B\ddot{\theta} + C = 0; \quad D\ddot{x} + E\ddot{\theta} + F = 0 \quad (29)$$

Solving the system of equations gives us the final result:

$$\ddot{x} = \frac{FB - CE}{AE - DB} \quad (30)$$

$$\ddot{\theta} = -\frac{C + A\ddot{x}}{B} \quad (31)$$

By converting (20) and (27) into a system of linear equations, we are able to solve for $\ddot{x}$ and $\ddot{\theta}$ in terms of the other known variables.

## 3. TUNING A PID CONTROLLER

Equation (1) discusses how the control action u(t) is calculated using a PID. The error e(t) is the error in the angle. To counteract the error in angle, a force is applied onto the cart. To create an efficient PID, the gain coefficients must be tuned. In order to prevent the PID output from being unrealistically large, the output u(t) is bounded within the range [-500, 500]. The initial conditions of the system are set to $(x, \theta) = (0, \frac{\pi}{4})$.

The proportional gain Kp is used to bring the current state closer to the target state. The integral gain Ki is used to eliminate steady-state error, which is especially important because of friction. The derivative gain Kd is used to decrease overshoot and undershoot.

### 3.1. Manual Tuning

Firstly, we adjust the proportional gain Kp until the PID is able to "balance" the rod with some oscillation. Some systems can be controlled only using a P controller, but this can't be done for an inverted pendulum on a cart due to the extreme instability (Figure 2).

In many occasions Ki is tuned before Kd. Tuning Kd first is the better approach here because it helps lower the oscillation (Figure 3), which allows us to adjust Ki later on to decrease steady-state error more accurately.

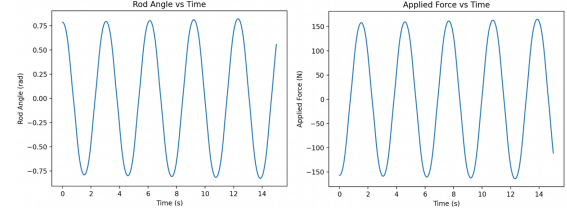

Figure 2. P Controller: (Kp, Ki, Kd) = (-200, 0, 0). Rod angle oscillates around the target angle of 0

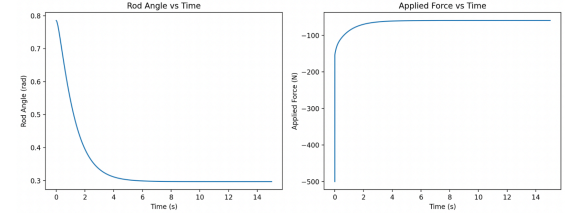

Figure 3. PD Controller: (Kp, Ki, Kd) = (-200, 0, -100). Minimal oscillation but there is steady state error.

Finally, Ki can be adjusted to eliminate the steady-state error (Figure 4).

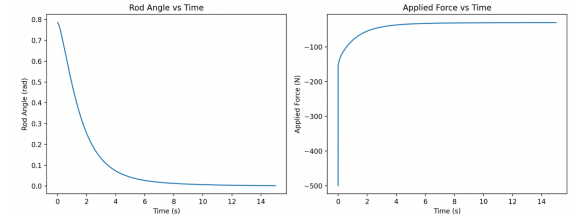

Figure 4. PID Controller: (Kp, Ki, Kd) = (-200, -20, -100). Minimal oscillation with no steady-state error.

Manual tuning allows us to generate a PID with decent but not optimal control. It's one of the most common ways of tuning PIDs, but there are definitely better options, especially in simulation. We can use optimization techniques to search for better gain coefficients.

### 3.2. Optimization-Based Tuning

In order to tune a PID using optimization, we must create a function where the parameters are the gain coefficients such that when the function is minimized, the optimal gain coefficient values have been found. First, we run the simulation for 15 seconds (15000 iterations) using a PID with the objective function parameters as the gain coefficients. The relatively long run duration of 15 seconds helps ensure the PID can balance the rod in the long-term. The objective function is the following:

$$f(K_p, K_i, K_d) = \sqrt{\frac{1}{n} \sum e(t)^2}$$

$$+0.0001(|K_p| + |K_i| + |K_d|) \tag{32}$$

where n is the number of iterations. The optimizer gradually tweaks the function parameters (gain coefficients) until a local minimum is found. The function is the sum of the root mean squared error (RMSE) of the rod angle and the absolute sum of the gain coefficients. We add the gain coefficients because it allows the optimizer to find control methods that require less force. A small constant is multiplied to the gain sum in order to prevent the optimizer from prioritizing low gains over low error.

To optimize the objective function, we use scipy.optimize.minimize. It's important to choose a good initial condition to help avoid undesired local minimums, so we experiment with various initial conditions to find the best. The following PID was obtained with an initial condition of (Kp, Ki, Kd) = (-300, 0, -100) (Figure 5).

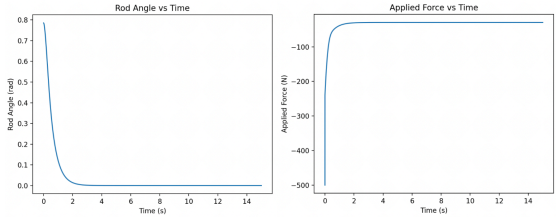

Figure 5. Optimized PID (Kp, Ki, Kd) = (-308.08, -63.55, -94.96) the error quickly decreases to zero and stays at 0.

The optimized PID also performs well at different starting rod angles, such as $\theta = \pi/6$ instead of $\theta = \pi/4$ (Figure 6).

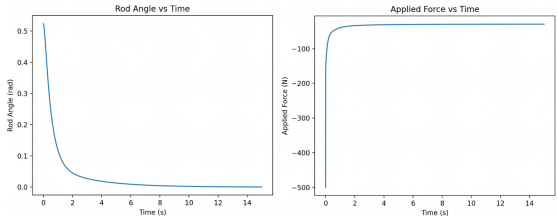

Figure 6. Optimized PID starting at $\theta = \pi/6$

Alternatively, we can use mean absolute error (MAE) instead of RMSE in our objective function. Using MAE makes convergence slower because the gradient curve of RMSE tends to be more differentiable and "smooth" than MAE. On a 2017 MacBook Pro, using RMSE allows for convergence in 167.023s, while MAE takes 236.14s. The overall PID performance is similar to that of using RMSE (Figure 7).

## 4. CONCLUSION

Using Lagrangian Mechanics, we successfully modeled the motion of an inverted pendulum on a cart when forces are applied onto the cart. This allowed us to simulate the system and develop PID controllers for the system. We used both manual tuning and optimization-based tuning methods. For optimization-based tuning, we can use PID coefficients similar to those yielding positive results from manual tuning as the initial conditions. This facilitates the generation of highly optimized PID controllers.

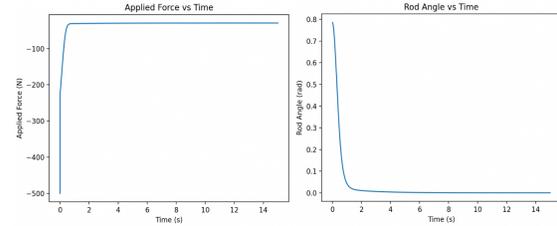

Figure 7. Optimized PID using MAE: (Kp, Ki, Kd) = (-289.57, -77.18, -60.65). The error approaches zero quicker than with RMSE but the error stays slightly above zero while slowly decreasing.

While we achieved positive results with a PID, there are many other options for control systems. For example, more sophisticated methods like linear quadratic regulator (LQR) and model predictive control (MPC) could have been used, which would allow for swing-up and swing-down of the rod [5]. Learning-based controllers such as imitation learning and reinforcement learning can also be tried. Some next steps would be comparing simulation physics with real-life physics. More accurate physics models can be derived with optimization or learning-based system identification techniques.